\documentclass[usenatbib]{emulateapj}
\usepackage{graphicx}
\usepackage[flushleft]{threeparttable}
\usepackage[usenames,dvipsnames,svgnames,table]{xcolor}
\usepackage{hyperref}
\definecolor{darkblue}{rgb}{0.0,0.0,0.3}
\hypersetup{colorlinks,breaklinks,
            linkcolor=darkblue,urlcolor=darkblue,
            anchorcolor=darkblue,citecolor=darkblue}
\usepackage{amsmath,amssymb}
\usepackage{amsmath}

\begin{document}

\def\etal{et al.\ \rm}
\def\ba{\begin{eqnarray}}
\def\ea{\end{eqnarray}}
\def\etal{et al.\ \rm}
\def\Fdw{F_{\rm dw}}
\def\Tex{T_{\rm ex}}
\def\Fdis{F_{\rm dw,dis}}
\def\Fnu{F_\nu}
\def\FJ{F_J}

%%%%%%%%%%%%%%%%%%%%%%%%%%%%%%%%%%%%%%%%%%%%%%%%%%%%%%%%%%%

\title{Accretion and Orbital Inspiral in Gas-Assisted Supermassive Black Hole Binary Mergers}

\author{Roman R. Rafikov\altaffilmark{1}}
\altaffiltext{1}{Institute for Advanced Study, 1 Einstein Drive, Princeton NJ 08540; 
rrr@ias.edu}

%%%%%%%%%%%%%%%%%%%%%%%%%%%%%%%%%%%%%%%%%%%%%%%%%%%%%%%%%%%

\begin{abstract}
Many galaxies are expected to harbor binary supermassive black holes (SMBHs) in their centers. Their interaction with the surrounding gas results in accretion and exchange of angular momentum via tidal torques, facilitating binary inspiral. Here we explore the non-trivial coupling between these two processes and analyze how the global properties of externally supplied circumbinary disks depend on the binary accretion rate. By formulating our results in terms of the angular momentum flux driven by internal stresses, we come up with a very simple classification of the possible global disk structures, which differ from the standard constant $\dot M$ accretion disk solution. Suppression of accretion by the binary tides, leading to a significant mass accumulation in the inner disk, accelerates binary inspiral. We show that once the disk region strongly perturbed by the viscously transmitted tidal torque exceeds the binary semi-major axis, the binary can merge in less than its mass-doubling time due to accretion. Thus, unlike the inspirals driven by stellar scattering, the gas-assisted merger can occur even if the binary is embedded in a relatively low mass disk (lower than its own mass). This is important for resolving the ``last parsec'' problem for SMBH binaries and understanding powerful gravitational wave sources in the Universe. We argue that the enhancement of accretion by the binary found in some recent simulations cannot persist for a long time and should not affect the long-term orbital inspiral. We also review the existing simulations of the SMBH binary-disk coupling and propose a numerical setup, which is particularly well suited for verifying our theoretical predictions.
\end{abstract}

\keywords{accretion, accretion disks --- galaxies: nuclei ---
(galaxies:) quasars: supermassive black holes}

%%%%%%%%%%%%%%%%%%%%%%%%%%%%%%%%%%%%%%%%%%%%%%%%%%%%%%%%%%%

%%%%%%%%%%%%%%%%%%%%%%%%%%%%%%%%%%%%%%%%%%%%%%%%%%%%%%%%%%%
%%%%%%%%%%%%%%%%%%%%%%%%%%%%%%%%%%%%%%%%%%%%%%%%%%%%%%%%%%%

\section{Introduction}  
\label{sect:intro}

%%%%%%%%%%%%%%%%%%%%%%%%%%%%%%%%%%%%%%%%%%%%%%%%%%%%%%%%%%%

Supermassive black hole (SMBH) binaries are expected to be a natural outcome of the hierarchical structure formation in the Universe \citep{Volonteri03,Volonteri09}. Mergers of galaxies implant their central SMBHs into the bulge of a merged galaxy, where stellar dynamical processes drive their inspiral into the center on Gyr time scales \citep{Begelman}. There SMBHs eventually form gravitationally bound binaries, which could be powerful sources of gravitational wave emission detectable with both pulsar timing arrays \citep{Lommen} and future space-based gravitational wave antennae \citep{Vitale}. 

Orbits of the SMBH binaries are expected to shrink, initially due to their gravitational interaction with the surrounding stars. However, at some point (typically at separations of $10^{-2}-1$ pc) the purely stellar dynamical processes are expected to become inefficient at driving the binary inspiral, at least in spherical haloes \citep{Yu,Vasiliev}. This gives rise to the so-called "last parsec problem" --- SMBH binaries stalling their orbital evolution and not being able to merge due to the gravitational wave emission within a Hubble time \citep{Yu,Milos}. This problem is not so obvious for SMBH binaries embedded in triaxial stellar haloes \citep{Vasiliev} but more work needs to be done to fully understand this issue. 

Quite naturally, SMBH binaries are also expected to interact with the gas that reaches the galactic center, the same gas that would fuel normal AGN activity of a single central black hole. Its accretion by the binary gives rise to a plethora of time-variable phenomena in the electromagnetic domain \citep{Tanaka,Ju}, allowing us to identify some candidate systems such as OJ 287 \citep{Valtonen} and PG 1302-102 \citep{Graham}. Moreover, gravitational coupling between the SMBH binary and the surrounding gas is expected to drive orbital evolution of the former, potentially alleviating the last parsec problem. Understanding this process  requires careful treatment of the exchange of mass and angular momentum between the circumbinary disk and the SMBH binary. Recent numerical results suggest that some details of this exchange are highly non-trivial (see \S \ref{sect:numerics}) and this motivates our study to a certain degree. 

The goal of this work is to place the orbital evolution of the SMBH binary in the context of the global viscous evolution of the circumbinary disk. Since both the binary inspiral and the degree to which disk accretion gets modified by the binary torque depend on the same process --- its angular momentum exchange with the disk, the two phenomena must be directly related. Understanding the form of this relation in different situations is the main purpose of this study.

Throughout this work we will also highlight the significance of the viscous angular momentum flux $\FJ$ (introduced in \S \ref{sect:struct}) for describing the evolution of the circumbinary disks. We show that using this variable instead of the more conventional surface density $\Sigma$ provides a useful diagnostics of the disk state and dramatically simplifies the classification of possible outcomes of the coupled binary-disk evolution. 

Even though in this work we will mainly talk about the disks around SMBH binaries, the majority of our results can be naturally translated to other similar astrophysical systems. One obvious example would be the protoplanetary disks orbiting young stellar binaries \citep{Rosenfeld,Czekala}. These disks are the birthplaces of circumbinary planets such as the ones recently discovered by the {\it Kepler} mission \citep{doyle_2011,welsh_2012}. Understanding their properties has been a subject of a number of recent studies \citep{Martin,Vartanyan}.

This work is structured as follows. After describing the basic setup in \S \ref{sect:main}, we explore the global evolution of the disk driven by its internal stresses and binary torque in \S \ref{sect:struct}, in particular the possible steady and quasi-steady outcomes. We describe the implications of these results for the binary inspiral in \S \ref{sect:orbital}, and then review the existing numerical studies of the same problem in \S \ref{sect:numerics}. Our discussion (\S \ref{sect:discussion}) includes the assessment of the validity of our assumptions (\S \ref{sect:validity}) and the description of the numerical setup optimized for exploring the long-term coupled evolution of the SMBH binary-disk system (\S \ref{sect:ideal}). 

%%%%%%%%%%%%%%%%%%%%%%%%%%%%%%%%%%%%%%%%%%%%%%%%%%%%%%%%%%%
%%%%%%%%%%%%%%%%%%%%%%%%%%%%%%%%%%%%%%%%%%%%%%%%%%%%%%%%%%%

\section{Basic setup.}  
\label{sect:main}

%%%%%%%%%%%%%%%%%%%%%%%%%%%%%%%%%%%%%%%%%%%%%%%%%%%%%%%%%%%

%%%%%%%%%%%%%%%%%%%%%%%%%%%
\begin{figure*}
\centering
\includegraphics[width=0.9\textwidth]{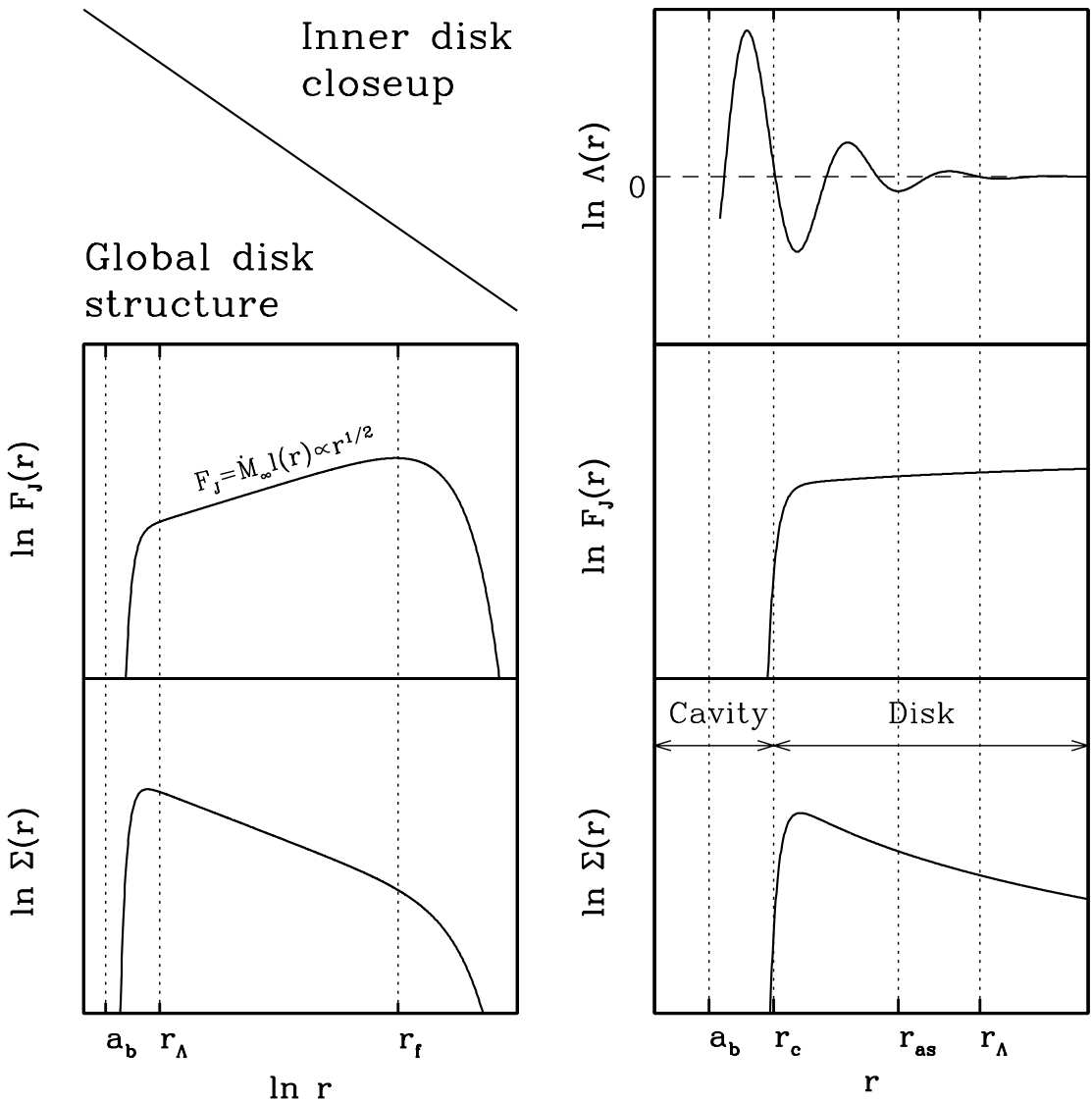}
\caption{Schematic illustration of the basic circumbinary disk properties, both global (left) and near the binary (right). Different panels show the behavior of the surface density $\Sigma$, angular momentum flux due to internal stresses $\FJ$, and binary torque density $\Lambda$. Mass is supplied at the large radius $r_f$, inward of which a constant-$\dot M$ disk develops due to internal stresses. Upon reaching the binary (with semi-major axis $a_b$) a cavity forms at $r=r_c$, within which $\Sigma$ and $\FJ$ are reduced dramatically. The disk can be considered as axisymmetric outside $r_{as}$. The binary torque (showing oscillatory radial behavior, see \S \ref{sect:tidal}) can be neglected outside the radius $r_\Lambda$. The global distribution of $\FJ\propto \dot M_\infty r^{1/2}$ reflects the initial state of the disk, before it has been perturbed by the binary torque. See text for details.
\label{fig:illustration}}
\end{figure*}
%%%%%%%%%%%%%%%%%%%%%%%%%%%

Following the majority of existing studies, we focus on a binary consisting of two black holes of mass $M_p$ (primary) and $M_s\le M_p$ (secondary), coplanar and corotating with the gaseous circumbinary disk. Binary semi-major axis is $a_b$, and it can have non-zero eccentricity $e_b$; its total mass is $M_b=M_p+M_s$, and mass ratio is $q\equiv M_s/M_p<1$. Misaligned and counter-rotating binary-disk systems have been previously considered by \citet{Nixon} and \citet{ivanov_2015}.

We assume that gas arrives from afar and circularizes into a disk-like configuration at some large distance $r_f\gg a_b$ from the barycenter of the binary. This is the radius at which the feeding of the disk with gas is determined. Starting at $r_f$ the initial ring-like configuration will viscously spread inwards (and outwards), with no external angular momentum injection until it reaches the binary. During this stage, viscous stresses drive the disk towards the state with {\it radially constant} mass accretion rate $\dot M$ and no torque at the center. Thus, the natural initial condition for the circumbinary disk just starting to interact with the SMBH binary should be a standard constant $\dot M$ disk first studied by \citet{shakura_1973}. 

The value of $\dot M$ is determined by the rate at which gas is added to the disk at $r_f$. For simplicity, we will assume continuous mass supply at $r_f$ at a steady rate $\dot M_\infty$ starting at some moment of time $t=0$ (but it should be remembered that this assumption is not crucial form our analysis). 

As the inner edge of the disk approaches $a_b$, gravitational effect of the non-axisymmetric component of the binary potential becomes important. It excites density waves, which propagate away and ultimately dissipate in the disk not too far from the center, imparting positive angular momentum into the disk fluid. As a result, the gas inflow is expected to slow down \citep{artymowicz_1994}, leading to the formation of the inner cavity in the disk with a characteristic radius $r_c$, see Figure \ref{fig:illustration} for illustration. A number of numerical studies \citep{MacFadyen2008} find $r_c\approx 2a_b$ for binaries with comparable components ($q\sim 1$), with some dependence on the binary eccentricity $e_b$ \citep{Pelupessy}. The suppression of accretion by the cavity is likely not perfect \citep{artymowicz_1996}, as the gas can flow into it via localized streams letting the binary to accrete at some finite rate $\dot M_b$, as discussed in \S \ref{sect:numerics}. 

Simulations also find that close to the edge of and inside the cavity ($r\lesssim r_c$) gas flow can hardly be considered as azimuthally uniform, see \S \ref{sect:numerics}. They clearly show  highly non-azimuthal motions in the form of dense gas streams inside the cavity. Also, the large scale spiral density waves launched by the binary result in significant radial motions at $r\approx r_c$. Moreover, simulations often exhibit the development of an eccentric instability in the disk resulting in the elliptical shape of the inner disk cavity \citep{MacFadyen2008,Cuadra2009}. 

However, already at $r\gtrsim$ several$\times a_b$ radial velocities induced by the binary torques become much smaller than the local Keplerian speed. Density waves could still be present in this part of the disk but the radial perturbations they induce on the gas motion is going to be small. Also, they perturb the disk only during the  short intervals when the wave passes across a particular fluid element. As a result, we can consider this part of the disk as axisymmetric, at least in the time-averaged sense, starting at some radius $r_{as}\gtrsim r_c$ (see Figure \ref{fig:illustration}). This is the disk region that we will focus on in this study.

%%%%%%%%%%%%%%%%%%%%%%%%%%%%%%%%%%%%%%%%%%%%%%%%%%%%%%%%%%%
%%%%%%%%%%%%%%%%%%%%%%%%%%%%%%%%%%%%%%%%%%%%%%%%%%%%%%%%%%%

\section{Global structure of the disk}  
\label{sect:struct}

%%%%%%%%%%%%%%%%%%%%%%%%%%%%%%%%%%%%%%%%%%%%%%%%%%%%%%%%%%%

In the axisymmetric zone, at $r\gtrsim r_{as}$, one can characterize the disk via the azimuthally averaged surface density $\Sigma(r)$. Its evolution is governed by the well-known equation following from the mass and angular momentum conservation \citep{lynden-bell_1974,papaloizou_1995}
\ba
\frac{\partial \Sigma}{\partial t} & = & \frac{1}{2\pi r}
\frac{\partial \dot M}{\partial r},
\label{eq:evSigma}\\
\dot M & = & \left(\frac{d l}
{d r}\right)^{-1}\frac{\partial T_{r\phi}}{\partial r}
+2\frac{\Sigma\Lambda}{\Omega}.
\label{eq:dotM}
\ea
Here $\dot M(r)$ is the local value of the mass accretion rate (defined to be positive for {\it inflow}), $T_{r\phi}$ is angular momentum flux due to the $r$-$\phi$ component of the {\it internal stress} in the disk, $\Lambda(r)$ is the specific (per unit mass) rate of {\it external} angular momentum injection, $\Omega(r)\approx (GM_b/a_b^3)^{1/2}$ is the angular frequency, which is close to Keplerian value at $r\gtrsim$ several$\times a_b$, and $l\equiv \Omega r^2$ is the specific angular momentum. 

Provided that stress is effected by some form of effective viscosity $\nu$, $T_{r\phi}$ is given by the {\it viscous} angular momentum flux $F_J$ \citep{Rafikov2013}
\ba
F_J\equiv -2\pi\nu\Sigma \frac{d\ln\Omega}{d\ln r}l=3\pi\alpha c_s^2\Sigma r^2,
\label{eq:F_J}
\ea
where $\nu$ is the kinematic viscosity expressed through the dimensionless parameter $\alpha$ and gas sound speed $c_s$ as \citep{shakura_1973}
\ba
\nu=\alpha \Omega ^{-1}c_s^2. 
\label{eq:nu}
\ea
Substituting $\FJ$ for $T_{r\phi}$ in equation (\ref{eq:dotM}) one arrives at the conventional form of the viscous evolution equation \citep{papaloizou_1995}. 

In general, the reduction of $T_{r\phi}$ to $\FJ$ is not guaranteed as the stress may be non-local (if it is provided by the disk self-gravity, see \citet{Balbus}) or anisotropic, such as that driven by the magneto-rotational instability (MRI). Nevertheless, in this work we will often resort to the ansatz (\ref{eq:F_J}) because it provides a simple illustration of the disk behavior and yields useful analytical results. Also, a number of numerical studies have been carried out assuming $\alpha$-viscosity \citep{DOrazio2013,Farris2014} and our adoption of the anzatz (\ref{eq:F_J}) allows a meaningful comparison with their results to be made. Thus, in the following we take $T_{r\phi}\to \FJ$ unless mentioned otherwise.

To provide full description of the disk properties one must specify the behavior of the angular momentum source term $\Lambda(r)$ due to the binary torque. We will address the details of the radial dependence of $\Lambda$ in \S \ref{sect:tidal}, but for now it is important to keep in mind that $\Lambda$ rapidly decreases with $r$. This behavior is supported by the numerical calculations and implies that, in practice, the term proportional to $\Lambda$ may be neglected in equation (\ref{eq:dotM}) outside some radius $r_\Lambda\gtrsim r_c$. There is a certain degree of freedom in the choice of $r_\Lambda$ (one can define it as a radius interior to which some fixed fraction, e.g. 90$\%$, of the total binary torque has been injected into the disk), although one expects $r_{as}\lesssim r_\Lambda$, see Figure \ref{fig:illustration}. However, this uncertainly does not affect the large-scale behavior of the disk. 

Outside $r_\Lambda$ disk evolution is driven only by internal stresses and we can drop $\Lambda$ term in equation (\ref{eq:dotM}). Adopting $l$ as an independent variable instead of $r$ equations (\ref{eq:evSigma})-(\ref{eq:dotM}) reduce to  
\ba
\frac{\partial \Sigma}{\partial t}  =  \frac{1}{2\pi r}\frac{\partial l}{\partial r}
\frac{\partial^2 \FJ}{\partial l^2},
\label{eq:evSigma1}
\ea
with the mass accretion rate given simply by \citep{Rafikov2013}
\ba
\dot M=\frac{\partial F_J}{\partial l}.
\label{eq:dotM1}
\ea

Under the reasonable assumption that disk mass contained within $r_\Lambda$ is small compared to both $M_b$ and the mass in the outer disk regions, we can safely assume that 
\ba
\dot M(r\to r_\Lambda)\to \dot M_b,
\label{eq:innerBC}
\ea
i.e. that $\dot M$ at the inner edge of a purely viscously-evolving region is equal to the mass accretion rate of the binary, which can be measured directly in simulations. 

On the other hand, in the outer disk ($r\lesssim r_f$) we expect $\dot M(r)\to \dot M_\infty$. 

We now consider different possibilities for the inner disk structure and evolution depending on the relation between $\dot M_b$ and $\dot M_\infty$.

%%%%%%%%%%%%%%%%%%%%%%%%%%%%%%%%%%%%%%%%%%%%%%%%%%%%%%%%%%%

\subsection{Steady disks.}  
\label{sect:steady}

%%%%%%%%%%%%%%%%%%%%%%%%%%%%%%%%%%%%%%%%%%%%%%%%%%%%%%%%%%%

We start by considering disks that have reached a steady state. In this case mass conservation necessarily implies that $\dot M(r)=\dot M_b=\dot M_\infty$ (in the absence of disk winds), i.e. $\dot M(r)=const$. 

\begin{figure}
\centering
\includegraphics[width=0.5\textwidth]{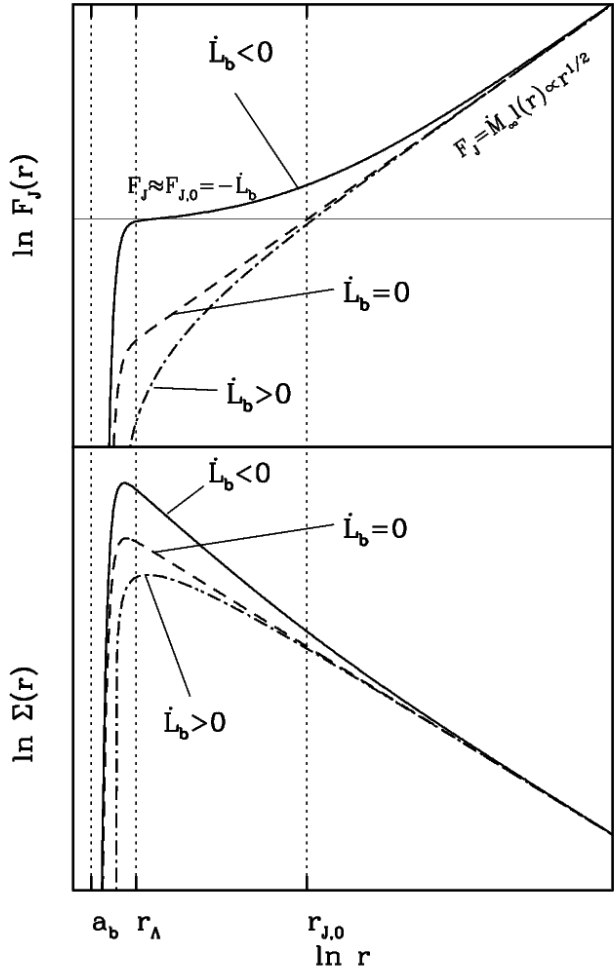}
\caption{Schematic representation of the profiles of the angular momentum flux due to internal stresses $\FJ$ and surface density $\Sigma$ in a {\it steady-state} disk with $\dot M_b=\dot M_\infty$. When the SMBH binary {\it loses} angular momentum to the disk ($\dot L_b<0$, \S \ref{sect:pos}, solid curves) gas piles up in the inner disk, compared to the case of $\dot L_b=0$ (\S \ref{sect:zero}, dashed curves); profile of $\FJ$ develops a plateau $\FJ\approx F_{J,0}=-\dot L_b$ near the binary, at radii $r\lesssim r_{J,0}$ (Eq. [\ref{eq:rJ0}]). When the SMBH binary {\it gains} angular momentum from the disk ($\dot L_b>0$, \S \ref{sect:neg}, dot-dashed curves), gas deficit must appear in the inner disk decoupling it from the binary torque and limiting the value of $\dot L_b$.
\label{fig:steady}}
\end{figure}

Setting left hand side of equation (\ref{eq:evSigma1}) to zero one immediately finds that in steady state
\ba
\FJ=T_{r\phi}=\dot M_\infty l +F_{J,0}
\label{eq:stst}
\ea
in the quasi-axisymmetric part of the disk, where $F_{J,0}$ is a constant. The factor multiplying $l$ follows from equation (\ref{eq:dotM1}). Such solutions are illustrated in Figure \ref{fig:steady}. A remarkable feature of this simple solution is that it does not involve the knowledge of the disk thermodynamic properties. However, according to equations (\ref{eq:F_J})-(\ref{eq:nu}), one does need to know $c_s(r)$ and $\alpha$ to infer $\Sigma(r)$ given the $\FJ$ profile (\ref{eq:stst}).

The value of $F_{J,0}$ is set by the tidal coupling between the binary and the disk. Taking the limit $l\to 0$ one immediately identified $F_{J,0}$ with the stress at the disk center, i.e. the rate of the angular momentum injection by the binary into the disk: 
\ba
F_{J,0}=-\dot L_b,
\label{eq:Td}
\ea
where $\dot L_b$ is the rate at which the binary loses\footnote{We expect the binary orbit to evolve slower than the disk, see \S \ref{sect:unsteady}, \ref{sect:orbital}.} its angular momentum.

Another way to understand equations (\ref{eq:stst}) and (\ref{eq:Td}) is to note that a full flux of angular momentum across a given radius is a sum of the outward viscous angular momentum flux $\FJ$ and the inward advective angular momentum inflow $-\dot M(r)l(r)$. In steady state, it must be equal to the binary angular momentum loss $-\dot L_b$ at every radius; also, $\dot M(r)=\dot M_\infty$. Thus,  $-\dot L_b=\FJ(r)-\dot M_\infty l(r)=F_{J,0}$, in agreement with equations (\ref{eq:stst}) and  (\ref{eq:Td}).

In steady state the value of $F_{J,0}$ (and $\dot L_b$) is determined by the details of the processes happening near the inner edge of the disk --- dependence of the stress on $\Sigma$, radial distribution of the binary torque density, and so on \citep{Kocsis1}, which are not known well. For that reason we chose to consider $F_{J,0}$ as a free parameter and explore implications of its variation for the disk-binary coupling. One can distinguish three possibilities for the disk structure depending on the value of $F_{J,0}$ (or $\dot L_b$), which we consider next.

%%%%%%%%%%%%%%%%%%%%%%%%%%%
%%%%%%%%%%%%%%%%%%%%%%%%%%%

%%%%%%%%%%%%%%%%%%%%%%%%%%%%%%%%%%%%%%%%%%%%%%%%%%%%%%%%%%%

\subsubsection{Steady disks: $F_{J,0}=\dot L_b=0$.}  
\label{sect:zero}

When the disk does not lose or gain angular momentum to the binary ($F_{J,0}=-\dot L_b=0$) via the combination of accretion and gravitational torque, its structure reduces to that of a conventional constant-$\dot M$ disk \citep{shakura_1973} with 
\ba
\FJ=\dot M_\infty l,~~~\dot M_\infty=3\pi\nu\Sigma.
\label{eq:constMdot}
\ea
The relation between $\Sigma$ and $\dot M_\infty$ follows from the definition (\ref{eq:F_J}). In this case all characteristics of the circumbinary disk outside $r_\Lambda$ --- radial distributions of $\Sigma$, disk temperature, spectral energy distribution, etc., should be the same as in the standard constant $\dot M$ disk with $\dot M=\dot M_\infty$ orbiting a {\it single} central object of mass $M_b$. 

The solution with $F_{J,0}=0$ provides a useful reference, to which other possible disk structures should be compared, see Figure \ref{fig:steady}. In particular, it describes the disk properties at the initial phases of its evolution, before its inner edge reaches the binary, see Figure \ref{fig:illustration} and \S \ref{sect:main}. It also represents a solution to which the disk structure converges far from the binary, see Figure \ref{fig:steady}.

%%%%%%%%%%%%%%%%%%%%%%%%%%%%%%%%%%%%%%%%%%%%%%%%%%%%%%%%%%%

\subsubsection{Steady disks: $F_{J,0}>0$, $\dot L_b<0$.}  
\label{sect:pos}

When the disk {\it absorbs} angular momentum from the binary($F_{J,0}=-\dot L_b>0$), the radial inflow of gas should slow down and it would accumulate in the inner disk, compared to the constant-$\dot M$ solution. The amount of mass pileup is determined by the degree to which $\Sigma$ must be increased to ensure $\dot M(r)=\dot M_\infty$ at the reduced (compared to constant $\dot M$ case) radial speed of the disk fluid.

We can rewrite the solution (\ref{eq:stst}) as 
\ba
\FJ=T_{r\phi}=\dot M_\infty \left(l +l_{J,0}\right),~~~
l_{J,0}\equiv -\frac{\dot L_b}{\dot M_\infty}.
\label{eq:stst1}
\ea
Constant $l_{J,0}$ has a simple physical interpretation in a disk that gains angular momentum: it is equal to the value of $l$ at the radius $r_{J,0}$ where the viscous angular momentum flux $\FJ$ in a constant-$\dot M_\infty$ accretion disk becomes equal to $-\dot L_b$, see \S \ref{sect:zero}. In other words,
\ba
r_{J,0}=\frac{l_{J,0}^2}{GM_b}=\left(\frac{\dot L_b}{\dot M_\infty}\right)^2\left(GM_b\right)^{-1},
\label{eq:rJ0}
\ea
as shown in Figure \ref{fig:steady}. The excess of $\Sigma(r)$ and $\FJ(r)$ over their values in a standard constant $\dot M$ disk is significant for $r\lesssim r_{J,0}$. Outside $r_{J,0}$ one finds $\FJ\to \dot M_\infty l$, see \S \ref{sect:zero}.

Constant $\dot M$ circumbinary disk solutions with $F_{J,0}>0$ and mass pileup have been previously studied by \citet{Shapiro}, \citet{Kocsis1,Kocsis2}, \citet{Liu} for different outer boundary conditions and disk properties. In particular, the constant $F_{J,0}=-\dot L_b$ in equation (\ref{eq:stst1}) corresponds to the {\it tidal barrier} introduced in \citet{Kocsis1}.

%%%%%%%%%%%%%%%%%%%%%%%%%%%%%%%%%%%%%%%%%%%%%%%%%%%%%%%%%%%

\subsubsection{Steady disks: $F_{J,0}<0$, $\dot L_b>0$.}  
\label{sect:neg}

If the binary torque were to inject {\it negative} angular momentum into the disk, $F_{J,0}=-\dot L_b<0$, the mass inflow towards the binary would be {\it accelerated} compared to the disk with the same $\dot M$ and  $F_{J,0}=0$. This would {\it reduce} $\Sigma$ near the binary compared to the standard constant $\dot M=\dot M_\infty$ disk, in contrast to what was found in \S \ref{sect:pos}.

The meaning of $r_{J,0}$ is different in such a disk: it represent the radius where the stress vanishes and $F_J\to 0$. If $\FJ$ is due to shear viscosity then equation (\ref{eq:F_J}) implies that $\Sigma=0$ interior to that radius, see dot-dashed curve in Figure \ref{fig:steady}. It is obvious that this radius cannot be very different from $a_b$: if this were happening at $r\gg a_b$ then the tidal coupling between the disk and the binary would not be present in the first place, since $\Lambda$ rapidly decays with $r$. This consideration robustly constrains the range of positive (negative) values that $\dot L_b$ ($F_{J,0}$) can potentially assume in steady state (if situations with $F_{J,0}<0$ are possible at all): $|\dot L_b|\lesssim \dot M_\infty l(a_b)$.

%%%%%%%%%%%%%%%%%%%%%%%%%%%%%%%%%%%%%%%%%%%%%%%%%%%%%%%%%%%

\subsection{Unsteady disks.}  
\label{sect:unsteady}

%%%%%%%%%%%%%%%%%%%%%%%%%%%%%%%%%%%%%%%%%%%%%%%%%%%%%%%%%%%

Whenever the accretion rate of the binary $\dot M_b$ is not matched to the external supply rate $\dot M_\infty$, mass accumulation (or loss) must occur in the disk. As a result, starting at $t=0$ with the standard constant $\dot M$ structure (our usual assumption, see \S \ref{sect:main}) the disk will evolve. 

As these changes are triggered by the tidal coupling to the central binary, disk evolution away from the constant $\dot M=\dot M_\infty$ structure will proceed from the inside out. Since this evolution is driven by internal stresses ("effective viscosity"), it should propagate out to large radii at the rate dictated by the viscous evolution. This process occurs on the local viscous time 
\ba
t_\nu(r)\sim \frac{r^2}{\nu} & \approx & 10^5~\mbox{yr}~\frac{0.1}{\alpha}\left(\frac{M_b}{10^8M_\odot}\right)^{-1/2}
\nonumber\\
&\times &\left(\frac{r}{10^{-2}\mbox{pc}}\right)^{3/2}
\left(\frac{h/r}{10^{-2}}\right)^{-2},
\label{eq:tnu}
\ea
see equation (\ref{eq:nu}). Here $h\equiv c_s/\Omega$ is the vertical thickness of the disk. Note that in most cases $\nu$ is a function of $r$. Following \citet{Rafikov2013} and \citet{Vartanyan} we introduce the concept of the {\it radius of influence} $r_{\rm infl}(t)$ defined via the implicit relation 
\ba
t_\nu(r_{\rm infl})=t
\label{eq:rinfl}
\ea
as the radius, at which the viscous time $t_\nu(r_{\rm infl})$ is equal to the time elapsed from the start of disk evolution. One can think of $r_{\rm infl}(t)$ as the radius, out to which the information about the changing state of the inner disk has been communicated by viscous stresses by the time $t$. Clearly, $r_{\rm infl}(t)$ monotonically increases in time. One can also introduce the associated specific angular momentum $l_{\rm infl}(t)\equiv \left(G M_b r_{\rm infl}\right)^{1/2}$. 

Equation (\ref{eq:evSigma1}) can be recast in a particularly simple form by switching from $\Sigma$ to the viscous angular momentum flux $\FJ$ and from $r$ to the specific angular momentum $l$ \citep{lynden-bell_1974,filipov_1984,lyubarskij_1987,Rafikov2013}:
\ba
\frac{\partial}{\partial t} \left( \frac{F_{J}}{D_{J}} \right) =
\frac{\partial^2\FJ}{\partial l^2},
\label{eq:evF_simple}
\ea
where
\ba
D_{J}\equiv -\nu r^2\frac{d\Omega}{dr}\frac{dl}{dr}
\label{eq:D_J}
\ea
is the diffusion coefficient, which depends on $F_{J}$ if $\nu$ depends on $\Sigma$. 

%%%%%%%%%%%%%%%%%%%%%%%%%%%
\begin{figure}
\centering
\includegraphics[width=0.5\textwidth]{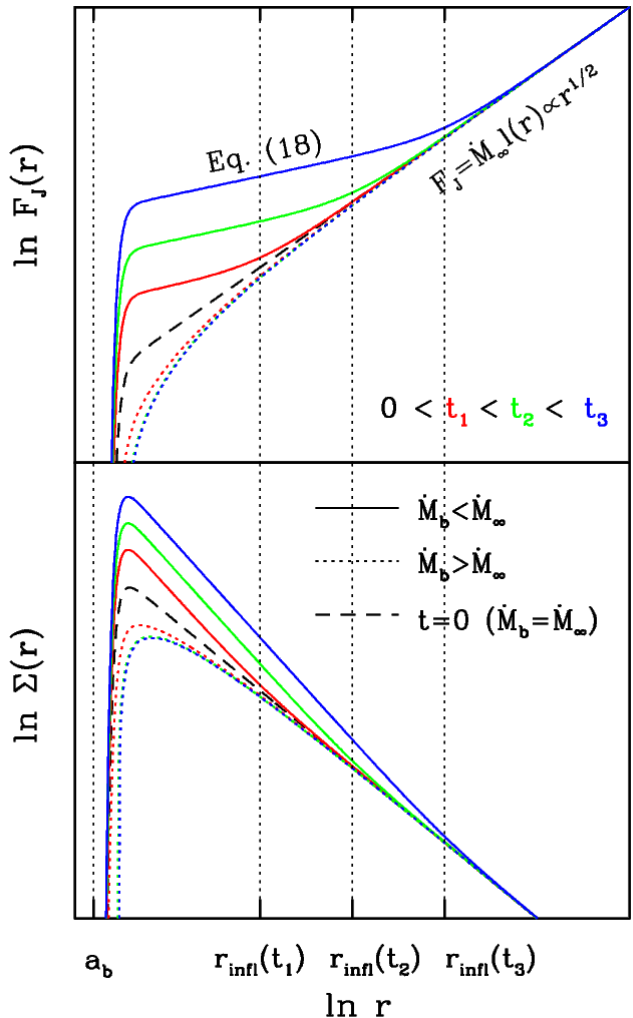}
\caption{Evolution of the disk properties in non-steady disks with $\dot M_b\neq\dot M_\infty$, i.e. when the binary accretion rate does not match the supply rate. Schematic representation of $\FJ$ (top) and $\Sigma$ (bottom) is shown at different moments of time, starting at $t=0$ (black dashed curve) when the disk state corresponds to that of a a standard constant $\dot M=\dot M_\infty$ disk. Different colors represent different moments of time as shown in Figure. Solid curves correspond to $\dot M_b<\dot M_\infty$, when the mass pileup develops in the inner disk and $\FJ$ distribution converges to that given by Eq. (\ref{eq:FJ_mod}) with the mass accretionj rate $\dot M_b$ for $r\lesssim r_{\rm inf}(t)$ at a given moment of time. Dotted curves correspond to $\dot M_b>\dot M_\infty$, a state which does not last since binary depletes the inner disk of mass and tidally decouples from it (illustrated by saturation of $\FJ$ and $\Sigma$ curves near the binary). 
\label{fig:nonsteady}}
\end{figure}
%%%%%%%%%%%%%%%%%%%%%%%%%%%

For power law scaling of $D_J$ with $l$ and $\FJ$ one can derive useful self-similar solutions of equation (\ref{eq:evF_simple}) with the inner boundary condition (\ref{eq:innerBC}) imposed at $l\to 0$. In the circumbinary disk setup\footnote{Self-similar viscous disk solutions in other astrophysical settings have been explored by e.g.  \citet{filipov_1984}, \citet{lyubarskij_1987}, \citet{Pringle}.} such solutions have been previously obtained by \citet{ivanov_1999} for a particular case of a non-accreting binary $\dot M_b=0$ and by \citet{Rafikov2013} for arbitrary (time-independent) $\dot M_b<\dot M_\infty$.

These solutions greatly help in understanding the qualitative features of the disk evolution. In particular, they show that, in agreement with qualitative expectations, the disk separates into two regions: outer, in which an unperturbed constant $\dot M_\infty$ solution persists, and inner, in which $\dot M=\dot M_b$ and $\FJ\to \dot M_b l + F_{J,0}$, with nonzero value of $F_{J,0}$. The boundary between them lies at the radius of influence $r_{\rm infl}$ and viscously expands. All disk properties evolve at the rate set by $t_\nu(r_{\rm infl})$, which coincides with the time since the system started evolving.

Even though the self-similar solutions are obtained for specific assumptions regarding the disk properties, namely the power law scaling of $\nu$ with $\Sigma$ and $r$ (or $D_J$ with $\FJ$ and $l$), their main features outlined above translate directly to the more general disk behavior. This is easy to understand by re-writing equation (\ref{eq:evF_simple}) (with the definition (\ref{eq:D_J})) as
\ba
\frac{\partial}{\partial t} \left(\FJ t_\nu\right) =\eta l^2
\frac{\partial^2\FJ}{\partial l^2},
\label{eq:evF_simple1}
\ea
where $\eta\equiv(-d\ln\Omega/d\ln r)(d\ln l/d\ln r)$ ($\eta=3/4$ for a Keplerian disk). According to this equation, at $r\gtrsim r_{\rm infl}$ the disk maintains it unperturbed structure with $\FJ=\dot M_\infty l$ because viscous stresses have not yet propagated out to the outer region (so $\partial/\partial t=0$ there). At $r\sim r_{\rm infl}$ both sides of the equation are important and internal stresses drive rearrangement on the disk characteristics away from the outer solution. In the inner disk, at  $r\lesssim r_{\rm infl}$ one naturally has $t_\nu\ll t_\nu(r_{\rm infl})$, so that the left hand side in equation (\ref{eq:evF_simple1}) can be neglected compared to the right hand side ($\partial/\partial t\sim \left[t_\nu(r_{\rm infl})\right]^{-1}$ since all disk properties including $\FJ$ evolve on characteristic timescale $t_\nu(r_{\rm infl})=t$). This, coupled with the boundary condition (\ref{eq:innerBC}) results in $\FJ\to \dot M_b l + F_{J,0}$ in the inner zone. However, unlike the situation covered in \S \ref{sect:steady}, now $F_{J,0}$ in the inner zone is not constant but {\it varies} on a characteristic timescale $t_\nu(r_{\rm infl})$. 

Matching of the inner and outer solution at $r\sim r_{\rm infl}$ implies that for $r\lesssim r_{\rm infl}$
\ba
\FJ(r,t)\approx \dot M_\infty l_{\rm infl}(t)+\dot M_b\left[l(r)-l_{\rm infl}(t)\right].
\label{eq:FJ_mod}
\ea
It has to be remembered that this solution is valid only as long as $\dot M_b$ is constant or varies on timescale longer than $t_\nu(r_{\rm infl})$. 

Angular momentum luminosity due to disk-binary coupling can be obtained from equation (\ref{eq:FJ_mod}) by taking the limit $l\to 0$ (or $r\to 0$), resulting in 
\ba
F_{J,0}=-\dot L_b\approx \left(\dot M_\infty -\dot M_b\right) l_{\rm infl}(t).
\label{eq:Tdnonsteady}
\ea
The inner disk solution (\ref{eq:FJ_mod}) reduces to a standard constant $\dot M_\infty$ solution with $F_{J,0}=0$ as $\dot M_b\to\dot M_\infty$.

%%%%%%%%%%%%%%%%%%%%%%%%%%%%%%%%%%%%%%%%%%%%%%%%%%%%%%%%%%%

\subsubsection{Unsteady disks: $\dot M_b<\dot M_\infty$.}  
\label{sect:normal}

Tidal binary-disk coupling can {\it suppress} $\dot M$ onto the binary compared to the case of a point mass, resulting in $\dot M_b<\dot M_\infty$. This expectation, motivated by 1D modelling \citep{ivanov_1999,armitage_2002,Kocsis1,Vartanyan} and early simulations \citep{artymowicz_1994,MacFadyen2008}, implies mass accumulation in the inner disk. This is similar to the emergence of the mass pileup found in \S \ref{sect:pos}. However, in the non-steady case the pileup grows in time. This general behavior is illustrated in Figure \ref{fig:nonsteady}. At any moment of time most of the mass accumulated in the inner region is located near the radius of influence $r_{\rm infl}$, as $\Sigma r^2\propto \FJ(r)/(\alpha c^2)$ increases with $r$, see definition (\ref{eq:F_J}). 

Whenever $\dot M_b<\dot M_\infty$, the inner solution has $F_{J,0}>0$, see equation (\ref{eq:Tdnonsteady}). This means that the disk necessarily gains angular momentum from the binary (i.e. $\dot L_b<0$) if the latter suppresses mass inflow in it. Moreover, in this case $F_{J,0}$ grows in amplitude as $r_{\rm infl}$ increases, unlike the situation covered in \S \ref{sect:steady}. For the non-accreting binary $\FJ$ in the inner disk becomes independent of $r$, see equation (\ref{eq:Tdnonsteady}) with $\dot M_b=0$ \citep{ivanov_1999}.

A scenario with $\dot M_b<\dot M_\infty$ provides a natural pathway to a steady state with $F_{J,0}>0$ (\S \ref{sect:pos}). Indeed, as the surface density at the inner disk edge increases, the gaseous streams flowing into the cavity will finally bring $\dot M_b$ in equilibrium with $\dot M_\infty$.

%%%%%%%%%%%%%%%%%%%%%%%%%%%%%%%%%%%%%%%%%%%%%%%%%%%%%%%%%%%

\subsubsection{Unsteady disks: $\dot M_b>\dot M_\infty$.}  
\label{sect:abnormal}

Motivated by the results of recent simulations (see \S \ref{sect:numerics} for in-depth discussion) we also comment on disk evolution in the case of binary accretion, {\it enhanced} as compared to the $\dot M_b=\dot M_\infty$ case. Whenever $\dot M_b>\dot M_\infty$ equation (\ref{eq:FJ_mod}) predicts that $F_{J,0}<0$, i.e. the disk {\it loses} angular momentum to the binary. This behavior, even if it occurs in the first place, cannot go on indefinitely: as the binary depletes the inner disk of mass, the tidal coupling between the two gets lost. As a result, $\dot M_b>\dot M_\infty$ cannot be maintained for an extended period of time. Another way to look at this outcome is to note that in this regime $\FJ\to 0$ at a finite radius (similar to the situation described in \S \ref{sect:neg}), and this radius grows with time. At some point this expansion  disconnects the binary from the disk. 

It is thus clear that situations with $\dot M_b<\dot M_\infty$ are rather pathological and cannot represent the long-term behavior of the circumbinary disk. Most likely the cavity edge will recede from the binary, reducing $\dot M_b$ to $\dot M_\infty$ so that the disk will rapidly (on time $\sim t_\nu(a_b)$) converge to a state described in \S \ref{sect:neg}.

%%%%%%%%%%%%%%%%%%%%%%%%%%%%%%%%%%%%%%%%%%%%%%%%%%%%%%%%%%%
%%%%%%%%%%%%%%%%%%%%%%%%%%%%%%%%%%%%%%%%%%%%%%%%%%%%%%%%%%%

\section{Orbital evolution of the binary}  
\label{sect:orbital}

%%%%%%%%%%%%%%%%%%%%%%%%%%%%%%%%%%%%%%%%%%%%%%%%%%%%%%%%%%%

Angular momentum of the binary 
\ba
L_b=\frac{q}{(1+q)^{2}}\left[GM_b^3 a_b(1-e_b^2)\right]^{1/2}
\label{eq:Lbin}
\ea
evolves as a result of the disk-binary coupling as $\dot L=-F_{J,0}$. This can be written as
\ba
t_a^{-1}\equiv\frac{\dot a_b}{a_b} = -2\frac{F_{J,0}}{L_b}-3\frac{\dot M_b}{M_b}-2\frac{\dot q}{q}\frac{1-q}{1+q}+\frac{2e_b\dot e_b}{1-e_b^2},
\label{eq:a_ev}
\ea
where $t_a$ is the characteristic orbital decay timescale. This expression shows that in general the semi-major axis of the binary evolves due to four effects: (1) loss (gain) of angular momentum to (from) the disk, (2) growth of the binary mass due to accretion, (3) variation of the mass ratio $q$ as a result of accretion, and (4) variation of the binary eccentricity. 

In our subsequent discussion we will neglect the evolution of $e_b$. Simulations \citep{Cuadra2009} often demonstrate initial increase of $e_b$ (starting from $e_b=0$) driven by the tidal disk-binary coupling, but then $e_b$ typically saturates at a value appreciably different from $1$ and does not evolve much past this point. This justifies our setting $\dot e_b\approx 0$ in equation (\ref{eq:a_ev}).

Accretion increases binary mass on a characteristic timescale $t_b\equiv M_b/\dot M_b=(\dot M_\infty/\dot M_b)t_M$, where we defined 
\ba
t_M\equiv\left(\frac{\dot M_\infty}{M_b}\right)^{-1}\approx
5\times 10^7~\mbox{yr}~\varepsilon_{0.1}\frac{\dot M_{\rm Edd}}{\dot M_\infty}.
\label{t_M}
\ea
Here $\dot M_{\rm Edd}$ is the Eddington mass accretion rate and $\varepsilon=0.1\varepsilon_{0.1}$ is the radiative efficiency of accretion.

Binary mass ratio evolves on a characteristic timescale $t_q$ defined via
\ba
t_q^{-1}\equiv\frac{\dot q}{q}=t_M^{-1}\frac{(1+q)^2}{q}\frac{\dot M_b}{\dot M_\infty}
\left(\frac{\dot M_s}{\dot M_b}-\frac{q}{1+q}\right),
\label{eq:t_q}
\ea
where $\dot M_s$ is the mass accretion rate of the secondary. 

This formula shows that if the secondary accretes at the rate, which is a higher fraction of $\dot M_b$ than its share of the binary mass (i.e. $\dot M_s/\dot M_b>M_s/M_b=q/(1+q)$) then $\dot q>0$ and $q$ tends to converge to unity. Simulations \citep{Farris2014,Young} suggest that this is indeed the case, with the secondary accreting disproportionately large fraction of $\dot M_b$ simply because it orbits closer to the inner edge of the disk and gets a better chance to intercept the material entering the cavity. \citet{Gerosa} fit the results of simulations by \citet{Farris2014} with a simple dependence $\dot M_s/\dot M_b\approx M_p/M_b=(1+q)^{-1}$. Plugging this into equation (\ref{eq:t_q}) one gets $\dot q=(1-q)^2/t_b$ so that $q\approx\tanh[(t+t_{q,0})/t_b]$, with $t_{q,0}\equiv (t_b/2)\ln[(1+q_0)/(1-q_0)]$ and $q_0$ being the mass ratio at $t=0$. 

This discussion demonstrates quite universally that, first, when $q\ll 1$, it grows on a short timescale $t_q\sim q t_b$. Second, $q$ approaches unity within $t\sim t_b$. Third, $t_q\to \infty$ as $q\to 1$ since $\dot M_s\to \dot M_b/2$ in this limit.

We now assess the contribution of the disk torque (first term in equation (\ref{eq:a_ev})) to the orbital evolution separately for steady (\S \ref{sect:steady}) and unsteady (\S \ref{sect:steady}) cases.

%%%%%%%%%%%%%%%%%%%%%%%%%%%%%%%%%%%%%%%%%%%%%%%%%%%%%%%%%%%

\subsection{Role of the disk torque: steady disks.}  
\label{sect:T_d_steady}

%%%%%%%%%%%%%%%%%%%%%%%%%%%%%%%%%%%%%%%%%%%%%%%%%%%%%%%%%%%

Effect of a steady disk with $\dot M_b=\dot M_\infty$ on the binary evolution depends on the sign of $\dot L_b$. 

When $\dot L_b=0$ binary conserves it angular momentum. Then, based on our previous discussion, we conclude that after $q$ reaches unity within $t\sim t_M$, the semi-major axis will keep shrinking only due to the growth of $M_b$. 

Situation is quite different if the binary injects positive angular momentum $F_{J,0}$ in the disk. Then, expressing $F_{J,0}$ via the characteristic radius $r_{J,0}$ defined by equation (\ref{eq:rJ0}), we can rewrite equation (\ref{eq:a_ev}) as
\ba
t_M\frac{\dot a_b}{a_b} &=& -\frac{2}{q}\frac{(1+q)^2}{\sqrt{1-e_b^2}}\left(\frac{r_{J,0}}{a_b}\right)^{1/2}
\nonumber\\
&-& 3-2\frac{t_M}{t_q}\frac{1-q}{1+q}.
\label{eq:a_ev_st}
\ea
with the three terms in the right-hand side describing the effects of the disk torque, growth of $M_b$ by accretion, and variation of $q$, respectively.

Starting from low $q_0$ it is clear that the evolution of $a_b$ is dominated by the disk torque and the increase of $q$, since both scale as $q^{-1}$ (see the discussion after equation [\ref{eq:t_q}]), unlike the second term describing the growth of $M_b$. However, as long as $r_{J,0}\gtrsim a_b$ the former dominates and disk torque shrinks the binary faster than both $q$ and $M_b$ evolve. 

If the mass ratio was close to unity from the start, $q_0\approx 1$, then its evolution would affect binary inspiral only weakly. Then the disk torque would again dominate $\dot a_b$ (over growth of $M_b$) as long as $r_{J,0}\gtrsim a_b$. 

In either case, the binary with a current semi-major axis $a_b$ and arbitrary $q$ will shrink and merge on a characteristic timescale 
\ba
t_m\approx t_M\left(\frac{a_b}{r_{J,0}}\right)^{1/2}\frac{q\sqrt{1-e_b^2}}{(1+q)^2}\lesssim t_M.
\label{eq:t_m_steady}
\ea
This estimate is obtained by integrating equation (\ref{eq:a_ev_st}) in which only the first term on the right hand side is retained, and assuming that $r_{J,0}$, as well as $q$ and $M_b$, remain constant during the inspiral (justified by the shortness of $t_m$).

Finally, in the unlikely case of significant $\dot L_b>0$ ($F_{J,0}<0$, see \S \ref{sect:neg}) binary would {\it drain} angular momentum from the disk, {\it slowing down} its inspiral. Whether negative $F_{J,0}=-\dot L_b$ can completely reverse the inspiral and drive expansion of the binary depends on the magnitude of $|\dot L_b|/(\dot  M_\infty l_b)$ ($l_b\equiv L_b/M_b$ is the specific angular momentum of the binary): if this ratio is $\gtrsim 3/2$ then such reversal is possible, see equation (\ref{eq:a_ev}). However, as noted in \S \ref{sect:neg} and confirmed by simulations \citep{ShiKrolik2012} (see \S \ref{sect:am_binary}) this ratio is unlikely to be large in magnitude since otherwise the binary would decouple from the disk.

%%%%%%%%%%%%%%%%%%%%%%%%%%%%%%%%%%%%%%%%%%%%%%%%%%%%%%%%%%%

\subsection{Role of the disk torque: unsteady disks.}  
\label{sect:T_d_unsteady}

%%%%%%%%%%%%%%%%%%%%%%%%%%%%%%%%%%%%%%%%%%%%%%%%%%%%%%%%%%%

In the case of an unsteady disk with $\dot M_b\neq\dot M_\infty$ (\S \ref{sect:unsteady}) we use equation (\ref{eq:Tdnonsteady}) to obtain
\ba
t_M\frac{\dot a_b}{a_b} &=& -\frac{2}{q}\frac{(1+q)^2}{\sqrt{1-e_b^2}}\left(1-\frac{\dot M_b}{\dot M_\infty}\right)\left[\frac{r_{\rm infl}(t)}{a_b}\right]^{1/2}
\nonumber\\
&-& 3\frac{\dot M_b}{\dot M_\infty} - 2\frac{t_M}{t_q}\frac{1-q}{1+q}
\label{eq:a_ev_unst}
\ea
instead of (\ref{eq:a_ev_st}). 

It is clear that the disk torque (first term in the right hand side) again dominates over other contributions for both low $q$ and $q\approx 1$, as long as $r_{\rm infl}\gtrsim a_b$. In this case, however, $r_{\rm infl}$ steadily increases with time, guaranteeing that beyond some point binary inspiral will be dominated by the disk torque. According to equation (\ref{eq:a_ev_unst}), this should happen when $r_{\rm infl}$ grows past $a_b$, which takes about a viscous time at $r=a_b$. According to equation (\ref{eq:tnu}) the latter is about $t_\nu(a_b)\sim 10^5$ yr for $a_b=10^{-2}$ pc, and this is much shorter than the time $t_M$, on which $M_b$ grows by accretion, see equation (\ref{t_M}).

Dropping the last two terms in equation (\ref{eq:a_ev_unst}), integrating it until the merger time $t_m$, and aproximating $\int_0^{t_m}\left[r_{\rm infl}(t^\prime)\right]^{1/2}dt^\prime\sim \left[r_{\rm infl}(t_m)\right]^{1/2}t_m$, we find the following implicit relation for $t_m$:
\ba
t_m\approx  t_M\left[\frac{a_b}{r_{\rm infl}(t_m)}\right]^{1/2}\left(1-\frac{\dot M_b}{\dot M_\infty}\right)^{-1}\frac{q\sqrt{1-e_b^2}}{(1+q)^2}.
\label{eq:t_m_unsteady}
\ea
The actual value of $t_m$ can be trivially determined from this formula once the time dependence of $r_{\rm infl}$ (set by the viscosity behavior) is known.

%%%%%%%%%%%%%%%%%%%%%%%%%%%%%%%%%%%%%%%%%%%%%%%%%%%%%%%%%%%
%%%%%%%%%%%%%%%%%%%%%%%%%%%%%%%%%%%%%%%%%%%%%%%%%%%%%%%%%%%

\section{Existing numerical results}  
\label{sect:numerics}

%%%%%%%%%%%%%%%%%%%%%%%%%%%%%%%%%%%%%%%%%%%%%%%%%%%%%%%%%%%

A number of numerical studies, starting with the pioneering work of \citet{artymowicz_1994}, addressed the coupled issues of the mass accretion by the binary, its tidal coupling to the disk, and the angular momentum budget. Below we briefly review the existing results in the context of our theoretical developments.

%%%%%%%%%%%%%%%%%%%%%%%%%%%%%%%%%%%%%%%%%%%%%%%%%%%%%%%%%%%

\subsection{Binary accretion rate}  
\label{sect:acc_rate}

%%%%%%%%%%%%%%%%%%%%%%%%%%%%%%%%%%%%%%%%%%%%%%%%%%%%%%%%%%%

It was first noted by \citet{artymowicz_1994} in their SPH simulations that the binary torques suppress accretion from the disk compared to the case of a single point mass. This conclusion has been confirmed by a more recent work of \citet{MacFadyen2008}, who found using {\it FLASH}, that the equal mass binary accretion rate is about $20\%$ of that in a single point mass case. Since then several other studies employing both SPH \citep{Cuadra2009} and grid-based \citep{ShiKrolik2012,DOrazio2013,Gold2014} techniques have arrived at a similar conclusion that the binary torque is efficient at suppressing (although not perfectly) the mass inflow from the disk. Some of these studies incorporated rather detailed physics, such as MHD effects, general relativity, disk self-gravity, and so on. These findings argue in favor of the scenario with $\dot M_b<\dot M_\infty$ presented in \S \ref{sect:normal} being the norm.

However, recently a different trend has been emerging. \citet{Farris2014} have observed $\dot M_b$ {\it exceeding} the rate at which a single black hole with the same total mass accretes from the same disk. Such an enhancement of accretion, unexpected both from the theoretical point of view and given the previous numerical results, was ascribed by \citet{Farris2014} to the complex nature of the binary-driven accretion inside the cavity, with gas penetrating the central hole via dense streams, splashing back onto the disk, etc. Revealing this complicated flow structure was made possible by using the moving-mesh code {\it DISCO} \citep{Duffell}, which allows treatment of the gas motion {\it inside} the cavity and around the individual black holes. Most previous grid-based studies of the same problem had to {\it excise} the central region enclosing the binary from the grid (with some exceptions \citep{Gunther,Hanawa,deval}, when non-polar grids were used) and could not follow gas motion inside the cavity as it was accreted by the binary components. However, this reasoning does not explain why such an  enhancement of accretion was not observed in previous SPH simulations, which also do not excise the central cavity. Nor does it explain the recent results of \citet{ShiKrolik2015}, who similarly to \citet{Farris2014} found an enhancement of accretion by the binary, but using a grid-based code (a modified version of {\it ZEUS} for MHD) with the excised central cavity. 

In any case, a situation with $\dot M_b>\dot M_\infty$ would argue in favor of the scenarios described in \S \ref{sect:abnormal}. However, as we argued there, this solution cannot represent a steady state as the rate at which mass is removed from the disk by the binary exceeds the rate at which gas is brought in from the outside. Over limited amount of time $\sim t_\nu(a_b)$ the inner disk should be depleted, reducing $\dot M_b$ and likely bringing it in equilibrium with $\dot M_\infty$. This, however, was not observed in simulations of \citet{Farris2014} and \citet{ShiKrolik2015}, probably because of their limited time span --- hundreds of binary periods, which is less than $t_\nu$ even close to the binary.

%%%%%%%%%%%%%%%%%%%%%%%%%%%%%%%%%%%%%%%%%%%%%%%%%%%%%%%%%%%

\subsection{Gravitational disk-binary coupling}  
\label{sect:tidal}

%%%%%%%%%%%%%%%%%%%%%%%%%%%%%%%%%%%%%%%%%%%%%%%%%%%%%%%%%%%

Manifestations of the tidal coupling between the binary and the disk are generally easier to interpret from simulations. A number of authors \citep{MacFadyen2008,Cuadra2009,ShiKrolik2012,Roedig2012,DOrazio2013} studied the radial dependence of the torque density $dT/dr\propto \Lambda(r)$ --- torque per unit radius exerted on the disk by the non-axisymmetric binary gravity (often additionally normalized by the disk surface density). It has been found quite invariably that $dT/dr$ exhibits {\it oscillatory} (in $r$) behavior near the binary, with the oscillation amplitude rapidly decaying with radius, so that effectively no torque is exerted on the disk  outside $\sim (3-4)a_b$. This general behavior, shown schematically in Figure \ref{fig:illustration} is robust with respect to the numerical scheme used (SPH or grid-based), and whether or not general relativity and/or MHD are included. 

It is very important to emphasize that the oscillatory radial profile of the torque density is very different from the scaling $dT/dr\propto |r-a_b|^{-4}$ normally adopted in 1D studies of the circumbinary disk evolution \citep{armitage_2002,Liu,Kocsis1}. Such scaling is motivated by the results of \citet{lin_1979a,lin_1979b} and  \citet{GT80} on $dT/dr$ behavior in the limit of a low-mass perturber, in the vicinity of its orbit, where the density of Lindblad resonances is high. Apparently, this prescription does not work well in circumbinary disks, where tidal coupling is dominated by a single (or a small number of) resonance for which the waveforms of the perturbed fluid variables, as well as $dT/dr$, exhibit oscillatory behavior, changing sign \citep{RP12,petrovich_2012}.

On the other hand, it also has to be remembered that $dT/dr$ measured in simulations represents the {\it excitation torque density} --- the rate at which angular momentum gets added by the binary tide to the density wave, which then propagates through the disk carrying this momentum and energy. Angular momentum gets passed to the disk only at the location where the wave dissipates, giving rise to a {\it deposition torque density} profile different from $dT/dr$. It is the deposition torque density that ultimately determines how the disk surface density evolves, see \citet{RP12}, \citet{petrovich_2012} and references therein. 

Total gravitational torque exerted on the disk (i.e. the integral of $dT/dr$ over the whole radial span of the disk) has also been evaluated in a number of studies. Its value may strongly depend on the details of the accretion flow near the binary. In particular, using MHD simulations \citet{ShiKrolik2012} found the integrated gravitational torque to exceed the value found by \citet{MacFadyen2008} in purely hydro simulations by more than an order of magnitude. This difference was ascribed to the dragging of more matter into the cavity by the MHD stresses in simulations of \citet{ShiKrolik2012}, although the gas motion within the cavity could not be followed (it was excised). This may have important implications for the overall angular momentum budget of the binary, see next.

%%%%%%%%%%%%%%%%%%%%%%%%%%%%%%%%%%%%%%%%%%%%%%%%%%%%%%%%%%%

\subsection{Binary angular momentum evolution}  
\label{sect:am_binary}

%%%%%%%%%%%%%%%%%%%%%%%%%%%%%%%%%%%%%%%%%%%%%%%%%%%%%%%%%%%

Total angular momentum loss of the binary $\dot L_b=-F_{J,0}$ is determined not only by the gravitational torque, but also by the angular momentum brought in by the accreted matter. While the former always decreases $L_b$, the latter always adds angular momentum to the binary (for disk corotating with the binary). \citet{MacFadyen2008} found the accreted angular momentum (per unit time) to be about $50\%$ of the gravitational torque, resulting in the net loss of the angular momentum by the binary, $\dot L_b<0$.

On the contrary, the MHD simulations of \citet{ShiKrolik2012} exhibit net {\it gain} of the angular momentum by the binary. This is a result of the angular momentum delivery by the accreting gas, which exceeds the gravitational torque on the disk, even though the latter is much higher than in simulations of \citet{MacFadyen2008}, see \S \ref{sect:tidal}. Nevertheless, the binary still shrinks in  \citet{ShiKrolik2012} even despite $\dot L_b>0$, because its net gain of the angular momentum gets offset by the increase of $M_b$. \citet{ShiKrolik2012} measure positive $\dot L_b\approx \dot M_\infty L_b/M_b$, which is not large enough to reverse the inspiral, see equation (\ref{eq:a_ev}) and the discussion in \S \ref{sect:T_d_steady}.

To summarize, we stress that understanding the long-term behavior of $\dot M_b/\dot M_\infty$ is very important. Solutions found in \S \ref{sect:struct} demonstrate that whether this ratio is above or below unity determines whether the binary gains or loses angular momentum, which then gets reflected in its orbital evolution (\S \ref{sect:orbital}). 

Unfortunately, the existing simulations are not well suited for resolving this issue since they are not well suited to attain a steady state. First, their boundary conditions are not designed for maintaining the flow of mass with constant $\dot M$ through the whole computational domain: the outer boundary is usually assumed to be absorbing. Second, their initial conditions often assume a disk of finite radial extent that freely expands both inwards and outwards. Third, it is also rare that a simulation (especially in 3D) is run for sufficiently long time, several viscous timescales at the outer edge of the computational domain. By design such simulations do not allow a steady state (or even a quasi-steady state with constant $\dot M_\infty$) to be established. In \S \ref{sect:ideal} we suggest a better simulation setup, which we believe should address these issues.

%%%%%%%%%%%%%%%%%%%%%%%%%%%%%%%%%%%%%%%%%%%%%%%%%%%%%%%%%%%
%%%%%%%%%%%%%%%%%%%%%%%%%%%%%%%%%%%%%%%%%%%%%%%%%%%%%%%%%%%

\section{Discussion.}  
\label{sect:discussion}

%%%%%%%%%%%%%%%%%%%%%%%%%%%%%%%%%%%%%%%%%%%%%%%%%%%%%%%%%%%

Theoretical developments presented in this work allow one to understand the long-term evolution of the circumbinary disks (\S \ref{sect:struct}) and their effect on orbital evolution of the central binary (\S \ref{sect:orbital}). In particular, we present a natural classification of the possible structures of an externally-fed disk in (quasi-)steady state. In general, we find that these disks differ considerably from the standard constant-$\dot M$ disk with no torque at the center \citep{shakura_1973}, in agreement with \citet{ivanov_1999} and \citet{Rafikov2013}.

Throughout this work we emphasize the utility of describing the disk structure in terms of the viscous angular momentum flux $\FJ$ (or total stress $T_{r\phi}$ in the case of MRI) at a given radius. Formulating our results in terms of $\FJ$ not only allows a simple classification of the possible outcomes, independent of the details of the disk thermodynamics. Also, the solutions themselves are extremely simple and can be easily connected to the evolution of the binary orbit. This has been previously demonstrated in \citet{Rafikov2013} who explored the SMBH binary-disk evolution assuming no binary accretion to occur ($\dot M_b\ll\dot M_\infty$). Our present study extends this work to arbitrary $\dot M_b$.

Tidal coupling to the disk can significantly accelerate orbital decay of the SMBH binary. The baseline for comparison is a standard constant-$\dot M$ solution (\ref{eq:constMdot}) with no mass pileup at the center. In this simple case $F_{J,0}$ is small, so that $r_{J,0}\lesssim a_b$ (or $r_{\rm infl}\lesssim a_b$). As a result, the binary inspiral is driven only by the growth of $M_b$ at roughly constant angular momentum $L_b$, and occurs on a characteristic timescale $t_M$. Neglecting the disk torque and specializing to the case of an equal mass binary ($q=1$, which is always reached within $\sim t_M$, see \S \ref{sect:orbital}) one finds $a_b\propto M_b^{-3}$. In other words, shrinking the binary orbit by a factor of 10 would require roughly doubling its mass and would take about $t_M$. 

Results of \S \ref{sect:T_d_steady} and \ref{sect:T_d_unsteady} demonstrate that the efficient tidal coupling to the disk can significantly accelerate the inspiral compared to the baseline case, if the inflowing gas piles up in the inner disk. The latter naturally occurs in steady state if the binary torque presents a substantial barrier to the gas inflow and $\Sigma$ in the inner disk goes up to ensure $\dot M_b=\dot M_\infty$ (\S \ref{sect:T_d_steady}). Tidal coupling speeds up the inspiral by $\sim (r_{J,0}/a_b)^{1/2}\gtrsim 1$ compared to $\dot L_b=0$ baseline, as shown by equation (\ref{eq:t_m_steady}). 

In the case of an unsteady disk with $\dot M_b<\dot M_\infty$ gas constantly accumulates in the inner disk and the size of the region significantly perturbed by the binary torque (compared to the constant-$\dot M$ solution [\ref{eq:constMdot}]) steadily grows. This expansion of $r_{\rm infl}$ explored in \S \ref{sect:unsteady} occurs on a relatively short viscous timescale $\sim t_\nu$, see equation (\ref{eq:tnu}). Even without the knowledge of the details of the $r_{\rm infl}(t)$ dependence, equation (\ref{eq:t_m_unsteady}) clearly demonstrates that the binary merger occurs within time $t_m\lesssim t_M$, as long as the pileup is significant and $r_{\rm infl}(t)\gtrsim a_b$. The expression for $t_m$ also shows that the merger takes less time for lower $\dot M_b/\dot M_\infty$. The merger is fastest for $\dot M_b\ll\dot M_\infty$, when $\FJ$ exhibits a plateau at $r\lesssim r_{J,0}$, see equation (\ref{eq:FJ_mod}) and \citet{Rafikov2013}.

Thus, tidal barrier allows binary orbit to shrink significantly in time short compared to its mass growth timescale $t_M$ (as well as $t_q$ on which binary mass ratio $q$ varies). This implies that both $M_b$ and $q$ would not change substantially while $a_b$ is reduced by order unity. Another important conclusion is that the circumbinary disk may be {\it considerably less massive} than the binary and still cause its significant orbital evolution.

This result may seem puzzling at first, but it has a simple nature. Internal stresses in the disk can transport angular momentum injected by the binary out to large distances. Given that the specific angular momentum $l$ grows as $r^{1/2}$, it takes only a small amount of mass at large $r$ to absorb the angular momentum of a massive SMBH binary (and disk mass grows with $r$ as well). This effect was described in \citet{Rafikov2013} for the non-accreting binary, but it should clearly be present also in the more general case studied in this work.

This picture is quite different from the scenario, in which the binary orbit is shrunk by the stellar dynamical processes, namely by gravitational scattering of stars closely approaching the binary. Then, to reduce its semi-major axis by a factor of two, the binary would need to scatter the mass in stars comparable to its own mass. This non-dissipative, collisionless process is thus rather inefficient compared to the tidal coupling to the disk, for which significant inspiral is possible even through interaction with a mass of gas $\lesssim M_b$, as we just showed.

We also make it pretty clear that the {\it enhanced} mass accretion by the binary, i.e. $\dot M_b> \dot M_\infty$ cannot persist for a long time and thus does not play a significant role in the orbital evolution of the SMBH binary. In this regard one should be careful to not over-interpret recent numerical claims of enhanced accretion by the binary \citep{Farris2014,ShiKrolik2015}, as they likely represent transient phenomena not relevant for the long-term evolution of the binary-disk system.

Our results on the significance of tidal coupling for the evolution of the SMBH binaries are in agreement with the findings of \citet{Vartanyan}, who explored the orbital evolution of the young stellar binary driven by the circumbinary protoplanetary disk. That study focused on the limit of a complete suppression of accretion ($\dot M_b\to 0$) and found that the tidal torque due to a massive circumbinary disk ($\sim 0.1M_\odot$) can easily bring a relatively compact binary with a period of $\sim 10$ d to a merger. Results of \S \ref{sect:orbital} should be relevant for extending this work to the case of non-zero $\dot M_b$.

In conclusion, we discuss the validity of the assumptions used in this work and describe some improvements that we recommend introducing in numerical simulations of the gaseous disks orbiting SMBH binaries.

%%%%%%%%%%%%%%%%%%%%%%%%%%%%%%%%%%%%%%%%%%%%%%%%%%%%%%%%%%%

\subsection{Validity of assumptions}  
\label{sect:validity}

%%%%%%%%%%%%%%%%%%%%%%%%%%%%%%%%%%%%%%%%%%%%%%%%%%%%%%%%%%%

Our discussion implicitly assumes internal stress in the disk to be produced by the shear viscosity. However, all the results would hold also if the stress were due to the MRI or some other mechanism of the angular momentum transport. In particular, in steady state one would still expect the solution (\ref{eq:stst}) to hold provided that $\FJ$ is replaced with $T_{r\phi}$ --- total stress at a given radius, including both magnetic and hydrodynamic (Reynolds) components. The important distinction of the MRI-driven transport is that the solution (\ref{eq:stst}) is expected to be maintained only in a {\it time-averaged} sense, i.e. after the fluctuations of the fluid variables caused by the MRI turbulence have been averaged out.

Throughout this work we assumed that the torque produced by the binary is injected in the disk on scales comparable to $a_b$, and is subsequently transported to larger distances purely by internal stresses. In principle, a different scenario may be possible, in which the density waves launched by the binary carry the angular momentum far out without substantial transfer to the disk fluid near the binary. In this case the binary would lose angular momentum and shrink, however, the inner disk would remain unaffected by the binary torque, and its structure would resemble that of the standard constant-$\dot M$ disk without angular momentum injection at the center. It might be tempting to interpret in favor of this scenario some recent numerical results on the disk-binary coupling, which find little perturbation to the disk structure by the binary torque (no significant pileup) and only weak suppression of the gas inflow (see more detailed discussion in \S \ref{sect:numerics}).

However, we find this interpretation implausible. \citet{R02} showed that a density wave propagating in the outer disk {\it always} evolves nonlinearly into a shock and dissipates, ultimately transferring all its angular momentum to the disk fluid, as long as $\Sigma$ and $T$ are decreasing functions of $r$. Thus, we should not expect the density waves to be able to transport the angular momentum injected by the binary very far from the inner edge of the disk. And if the disk ends up largely undisturbed by the presence of the binary on scales $\gtrsim a_b$, then this implies $\dot L_b\approx 0$.

%%%%%%%%%%%%%%%%%%%%%%%%%%%%%%%%%%%%%%%%%%%%%%%%%%%%%%%%%%%

\subsection{An "ideal" numerical setup}  
\label{sect:ideal}

%%%%%%%%%%%%%%%%%%%%%%%%%%%%%%%%%%%%%%%%%%%%%%%%%%%%%%%%%%%

As we discussed in \S \ref{sect:numerics}, existing simulations of circumbinary disks are not well suited for exploring the long-term, global evolution of the binary-disk system. Here we suggest a particular numerical setup for this task, which is designed keeping in mind analytical disk solutions described in \S \ref{sect:struct}. This setup has two important ingredients.

First, the initial state of the disk should closely correspond to its expected analytical steady state structure. Namely, outside several $a_b$ the disk should have the initial profile of its angular momentum flux correspond to the constant $\dot M$ disk, i.e. $\FJ(r)=\dot M_\infty l(r)$. According to the definition (\ref{eq:F_J}) this means that a certain profile of $\Sigma(r)$ must be set up in the beginning, depending on the radial behavior of $\alpha$ and disk temperature. 

Second, influx of mass at the rate $\dot M_\infty$ should be maintained at the outer boundary through the duration of the simulation. This can be relatively easy to implement if the mass inflow in the disk is driven explicitly by the shear viscosity \citep{Suzuki}, but may be more challenging in simulations with the angular momentum transport driven by the MRI. 

By having these two conditions fulfilled one would ensure that the outer regions of the disk do not significantly evolve, at least as long as the duration of the simulation is less than the viscous time at the outer boundary of the domain. It would also guarantee the steady mass supply towards the SMBH binary, exactly as expected in reality if the gas feeding radius $r_f$ significantly exceeds $a_b$. By measuring $\dot M_b$ one would directly determine its mismatch with the external mass supply rate $\dot M_\infty$. 

Over time binary torques and accretion will modify the inner disk structure. If $\dot M_b<\dot M_\infty$ for long enough time, a significant amount of mass will accumulate near the inner edge of the disk. Size of the region where mass piles up and $\FJ$ exceeds $\dot M_\infty l$ will expand in accordance with equation (\ref{eq:rinfl}). At some point $\dot M_b$ will equilibrate with $\dot M_\infty$, and the disk will reach a (quasi)-steady state. The value of $r_{\rm infl}$ at this time will effectively become $r_{J,0}$ given by equation (\ref{eq:rJ0}). The disk outside $r_{J,0}$ will maintain its original structure, except that $\FJ$ will be increased by $F_{J,0}\approx \dot M_\infty l(r_{J,0})$ on scales $r\lesssim r_{\rm infl}(t)$. By measuring the constant $F_{J,0}$ in $\FJ(l)$ at these radii one would directly obtain the rate at which the binary loses angular momentum to the disk. Note that with this approach such a measurement can be done far enough from the binary to avoid the complications related to the non-trivial flow structure inside the cavity. 
As mentioned before, it is unlikely that in steady state $F_{J,0}=-\dot L_b$ could be large and negative. Thus, even if $\dot M_b>\dot M_\infty$ initially, the system should rapidly relax to a state in which  $\dot M_b=\dot M_\infty$, after some mass has been evacuated from the disk center. The small negative value of $F_{J,0}\lesssim \dot M_\infty l_b$ can again be measured far from the binary, at $r\lesssim r_{\rm infl}(t)$.

The simplest numerical setup in which this evolution can be traced is a 2D hydrodynamical simulation using explicit viscosity \citep{MacFadyen2008,DOrazio2013,Farris2014}. Using this viscosity ansatz instead of MRI allows one to reduce dimensionality of the problem (from 3D to 2D) significantly speeding up the simulations. This is important since following viscous evolution of the disk on global scales requires rather long-term runs. Moreover, explicit viscosity allows the flow to have a more regular structure facilitating its diagnostics (measurement of $\dot M(r)$, $\FJ(r)$, etc.), compared to MRI, which typically requires long term averaging of the fluid variables. 

%%%%%%%%%%%%%%%%%%%%%%%%%%%%%%%%%%%%%%%%%%%%%%%%%%%%%%%%%%%
%%%%%%%%%%%%%%%%%%%%%%%%%%%%%%%%%%%%%%%%%%%%%%%%%%%%%%%%%%%

\section{Summary}  
\label{sect:sum}

%%%%%%%%%%%%%%%%%%%%%%%%%%%%%%%%%%%%%%%%%%%%%%%%%%%%%%%%%%%

We have explored the coupled evolution of the SMBH binary and its surrounding gaseous disk. By considering the global conservation of the angular momentum of the combined system and accounting for the viscous evolution of the disk we are able to classify the possible evolutionary outcomes into a handful of regimes. Using rather general analytical arguments, we arrive at the following conclusions.

\begin{itemize}

\item Evolutionary state of the disk can be most conveniently characterized in terms of the viscous angular momentum flux $\FJ$ (in the case of angular momentum transport driven by the effective shear viscosity), or total stress at a given radius, regardless of the thermodynamical properties of the disk. 

\item In steady state $\FJ$ is a {\it linear} function of the local specific angular momentum $l$ (Eq. [\ref{eq:stst}]). Its behavior is determined both by the externally imposed mass accretion rate through the disk $\dot M_\infty$ and the injection of the angular momentum by the binary. When the latter is non-negligible, the disk properties can differ significantly from those of a standard constant-$\dot M$ disk with the same $\dot M$.

\item Mass accumulation near the binary induced by the tidal torque leads to the loss of the binary angular momentum and shrinking of its orbit. This process is more efficient than the inspiral driven only by the increase of the binary mass due to gas accretion (at fixed angular momentum), provided that the disk region significantly perturbed by the binary torque ($r_{J,0}$ or $r_{\rm infl}$) exceeds its semi-major axis $a_b$.

\item Orbits of the SMBH binaries capable of producing gas pileup through their torque can shrink significantly even if the surrounding disk contains less mass than the binary itself. Orbital decay occurs on a timescale shorter than the time to double the binary mass by gas accretion. In this regard the gas-assisted inspirals are more efficient for solving the last parsec problem and giving rise to powerful gravitational wave sources than the inspirals driven by the stellar scattering alone.

\item SMBH binary can accrete from the disk at a rate higher than $\dot M$ of its single counterpart of the same mass (as suggested by some simulations) only for a short period of time. This transient state cannot affect its long-term orbital evolution.

\item Carefully designed simulations accounting for the expected (quasi-)steady state of the binary-disk system have the potential to verify these predictions and explore the long term evolution of both the SMBH binary and the disk.

\end{itemize}

Our results are applicable to other systems harboring circumbinary disks, e.g. young stellar binaries orbited by the protoplanetary disks \citep{Vartanyan}.

\acknowledgements

I am grateful to Diego Mu$\tilde {\rm n}$oz for useful comments on the manuscript. R.R.R. is an IBM Einstein Fellow at the IAS. Financial support for this study has been provided by NSF via grants AST-1409524,  AST-1515763, NASA via grant 14-ATP14-0059, and The Ambrose Monell Foundation.

%%%%%%%%%%%%%%%%%%%%%%%%%%%%%%%%%%%%%%%%%%%%%%%%%%%%%%%%%%%
%%%%%%%%%%%%%%%%%%%%%%%%%%%%%%%%%%%%%%%%%%%%%%%%%%%%%%%%%%%

\bibliographystyle{apj}
\bibliography{references}

\end{document}